# International Trade Finance from the Origins to the Present: Market Structures, Regulation and Governance

Olivier Accominotti and Stefano Ugolini

Abstract

This chapter presents a history of international trade finance – the oldest domain of international finance – from its emergence in the Middle Ages up to today. We describe how the structure and governance of the global trade finance market changed over time and how trade credit instruments evolved. Trade finance products initially consisted of idiosyncratic assets issued by local merchants and bankers. The financing of international trade then became increasingly centralized and credit instruments were standardized through the diffusion of the local standards of consecutive leading trading centres (Antwerp, Amsterdam, London). This process of market centralization/product standardization culminated in the nineteenth century when London became the global centre for international trade finance and the sterling bill of exchange emerged as the most widely used trade finance instrument. The structure of the trade finance market then evolved considerably following the First World War and disintegrated during the interwar de-globalization and Bretton Woods period. The reconstruction of global trade finance in the post-1970 period gave way to the decentralized market structure that prevails nowadays.

# 1. Introduction

Trade finance is the oldest domain of international finance. From the very beginnings of the history of international commerce, merchants and firms have been in need of working capital in order to finance their international commercial transactions and have looked for methods to reduce the risks involved in long-distance trade. Engaging in international trade requires immobilizing substantial amounts of capital for a relatively short (yet non-negligible) amount of time and also involves substantial risks for both exporters and importers. Trade finance encompasses all instruments and methods allowing firms to obtain capital in order to finance their international activities and/or reduce the risks involved in international trade. The modern literature distinguishes between the financing of international trade through products issued by banks (usually called "bank-intermediated trade finance") and the financing by the exporting and importing firms themselves (usually labelled "inter-firm trade credit").

Recently, renewed attention has been paid in the literature to the methods and instruments used in financing international trade (International Monetary Fund, 2011; Bank for International Settlements, 2014; Niepman and Schmidt-Eisenlohr, 2017). Researchers have highlighted how shocks to the supply of trade finance can have severe repercussions on the volume of world exports and stressed the role of such shocks in the "Great Trade Collapse" that followed the 2008 global financial crisis (Ahn, Amiti and Weinstein, 2011; Del Prete and Federico, 2014; Paravisini et al., 2015; and Niepman and Schmidt-Eisenlohr, 2016).

This chapter reviews the main developments on the global trade finance market from the emergence of the bill of exchange in the Middle Ages up to the present day. Our aim is to describe how the structure of this market evolved over time as well as the implications of these changes for its governance. We describe the trade credit instruments used at different times and the role played by banks in the main international financial centres in financing merchandise trade. Finally, we analyse how the main evolutions of the international monetary and financial system reshaped the structure and governance of international trade finance.

Historically, the most widespread instrument for financing merchandise trade was the bill of exchange, whose use became generalized in the medieval period. Bills of exchange are private written orders by one party to another to pay a given sum at a given date. As such a date is generally remote in the future, bills of exchange are not only means of national or international payment (as it is the case e.g. for cheques), but also credit instruments. In this chapter, we only focus on the use of bills as instruments of private credit for the financing of international trade. Initially, bills of exchange consisted of idiosyncratic assets which were issued by local merchant and banking firms. Trade finance products however became increasingly standardized from the sixteenth century onwards, and the financing of international commerce was progressively centralized around the successive leading trading centres of Antwerp, Amsterdam and London. This process of market centralization/product standardization culminated in the nineteenth century when London became the global centre for international trade finance and a very liquid market developed for sterling bills of exchange. At that time, a substantial share of global trade was financed through the London money market in which all types of investors participated. The structure of the global trade finance market however disintegrated in successive steps during the interwar and post-war years and its reconstruction in the post-1970 period gave way to a decentralized market where trade finance products are issued locally by banks in the exporters' and importers' countries.

The rest of the chapter is organized as follows. Section 2 provides a brief summary of the origins of the bill of exchange and its evolutions in the medieval and early modern periods. Section 3 describes the emergence of London as the global centre for trade finance in the nineteenth century. Section 4 explores how the market structure prevalent during the first globalization progressively disintegrated in the interwar period. Section 5 analyses how the trade finance market was reconstructed after the Second World War (WW2) and compares the methods and instruments used today in the financing of international trade with those in place before WW1. Section 6 concludes.

## 2. The emergence of trade finance products, 1100-1800

## 2.1. The origins of the bill of exchange in the medieval period

Trade finance was already present in the early civilizations of the Middle East (Joannès 2008) and was one of the pillars of the sophisticated banking systems developed in the Greek Mediterranean (Bogaert 1968; Cohen 1992). However, not much is known about the precise methods used to finance international trade in the Antiquity as well as during the age of the Arabic domination of the Mediterranean (Geva 2011). More information is available on the re-emergence of trade finance in the West after 1000 AD. From the perspective of international economic governance, the evolution of trade finance in Western Europe during the late medieval and early modern eras is an especially interesting case because it took place in a context of competing jurisdictions largely sharing common legal frameworks - especially thanks to the Roman heritage and canon law. Trade finance was also at the heart of discussions around the so-called "law merchant" or "*lex mercatoria*" – a set of commercial principles used by merchants across Europe.[1] It is therefore worth recalling how Europe ended up producing a well-defined international standard for the financing of international commerce, which pioneered later developments in other branches of international regulation.

De Roover (1953) describes the evolution of trade finance in Western Europe before the 19th century as marked by a number of gradual technical changes, the most important of which took place in the course of the 16th century. Until the 13th century, international trade was mainly financed on a local basis through fairly idiosyncratic lending practices. Commercial business was still organized as "caravan trade": missions were financed singularly through the creation of *ad hoc* partnerships (known in Italy as "*commendas*") among fellow citizens. Although trade flows often connected the entrepreneurs' place of origin with an international trading centre, their organisation and financing took place on a strictly bilateral foot. As a matter of fact, early commercial hubs like the popular Champagne fairs (12th-13th centuries) long lacked the clear juridical framework that would have allowed for the development of multilateral financial flows (Sgard 2015, p. 177).

---

[1] The concept of "*lex mercatoria*" has been popularized esp. by Malynes (1686) in opposition to common law (see below, section 3). According to Malynes (1686, p. 44), international "law merchant" was a branch of natural law mainly concerned with three domains: commodity trade, specie flows, and bills of exchange – defined respectively as the "body", "soul", and "spirit" of commerce. On the idea that the medieval "law merchant" is the basis of current international trade regulation, see e.g. Milgrom, North, and Weingast (1990) for a supportive view, and Volckart and Mangels (1999) for an opposite one; most of this recent historical discussion, however, does not actually address the question of the financing of international trade. From a juridical viewpoint, the current consensus is that "*lex mercatoria*" was actually not natural law, but a specific branch of common law (Baker 1979); it existed as such, until the 18th century, also in civil law countries (Cerutti 2003). To date, many jurists are however skeptical about the very concept of "law merchant" (see e.g. Kadens 2012).

**Figure 1. The medieval bill of exchange**

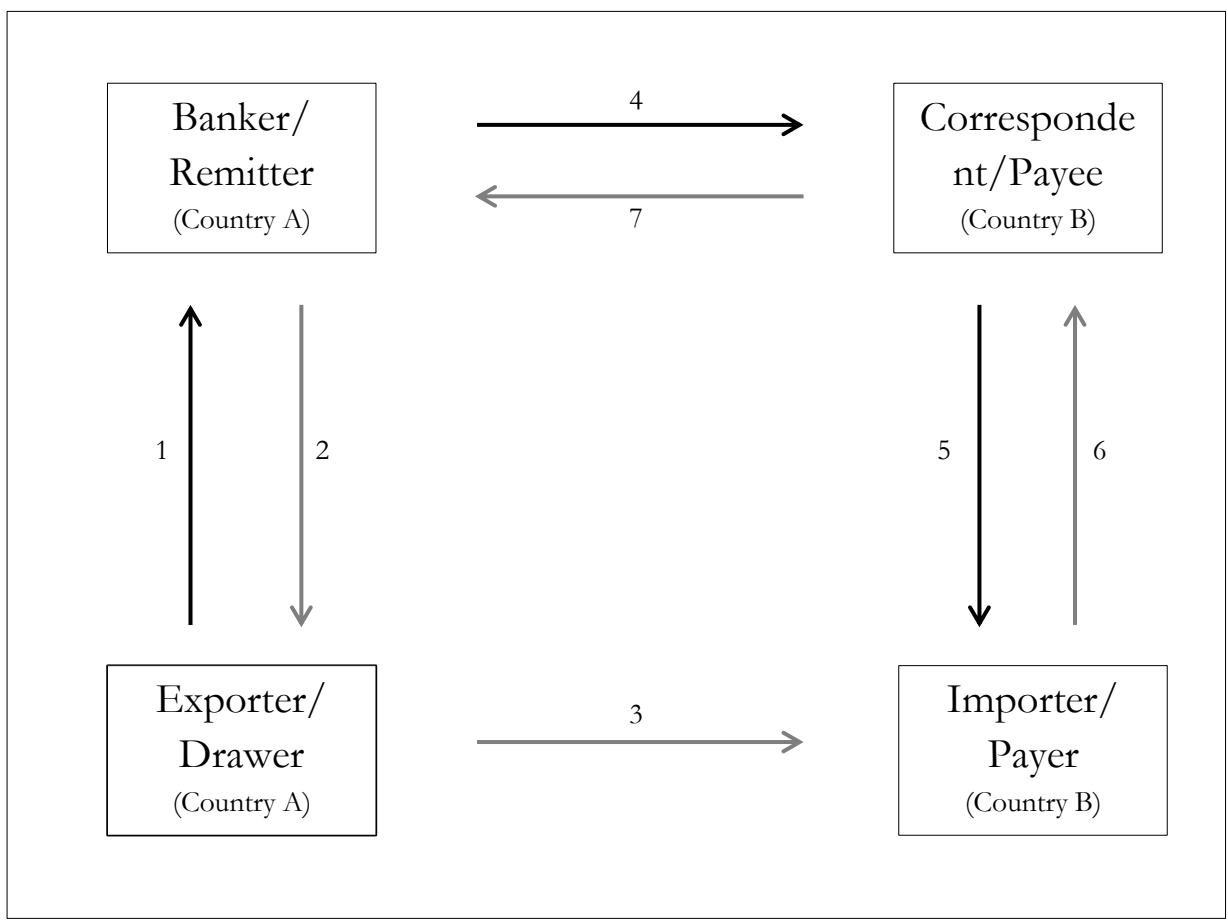

1. Issues bill of exchange; 2. Purchases the bill/provides cash; 3. Ships goods; 4. Sends the bill; 5. Presents the bill at maturity; 6. Pays the bill at maturity/provides cash; 7. Credits remitter at maturity.

This structure started to change during the 13th century when commercial business partially evolved into "sedentary trade", as merchants began to create stable networks of correspondents. Such a process, however, did not concern the whole of Europe; initially, it was only led by a number of Italian (especially, Tuscan) companies, and thus only covered the areas in which the latter were active – largely excluding, for instance, the German and Baltic regions, as well as all extra-European places except Constantinople (De Roover 1953, pp. 38-50). Within this new business model, trade flows were organized multilaterally among the different vertexes of each single correspondent network. Yet, within these new private trade networks, commercial business continued to be run by groups of fellow citizens (albeit on a much grander scale than before), and its funding came from within these very "clans". It was in this context that the future prime instrument of international trade finance – i.e., the *bill of exchange* – originally emerged. The original bill of exchange was not a standardized credit instrument, but rather a mere certificate – to be presented to a foreign correspondent – of a private credit contract passed between two *local* agents (De Roover 1953, pp. 91-94).

Figure 1 describes the functioning of the medieval bill of exchange. The example is drawn from De Roover (1953, pp. 45-47), who describes a transaction taking place in 1399-1400 between four Florentine businessmen, two based in the Flanders ("country A") and two based in Catalonia ("country B"). In order to finance his exports of Flemish goods to Catalonia, a Florentine merchant established in Bruges (the *drawer*) issued a bill of exchange and sold it to his Bruges-based banker (the *remitter*, also originally from Florence). By issuing the bill, the exporter ordered his importer (the *payer*, also a Florentine) to pay a given sum to a specified correspondent of the Bruges-based banker in Barcelona (the *payee*, another Florentine citizen). Through this operation, the exporter obtained cash from his Bruges banker before the importer paid for the goods and the banker therefore lent to the exporter on the security of the bill. After purchasing the bill, the Bruges banker sent it to his correspondent in Barcelona and, at maturity, the latter presented the bill of exchange for payment to the importer. Finally, once the bill had been encashed, the *payee* credited the *remitter*'s current account. In this example, the Bruges banker could secure repayment of his loan from the importer in Barcelona through his own correspondent in the city but, since the bill was not negotiable, he had to finance the commercial transaction with his own capital until the bill's maturity.

Medieval bills of exchange were not standardized securities freely exchangeable on an open market, but idiosyncratic loan contracts passed between members of a closed business network. One good example of the idiosyncratic nature of the financing of trade in the medieval period is a 15$^{th}$-century instrument known as "exchange on Venice" (*cambium ad Venetias*). As Venice became Europe's first stable commercial hub allowing for continuous trading throughout the year – or, as Luzzatto (1954) put it, a "permanent fair" – Florentine merchant and banking companies established branches there and started to play a substantial role in the funding of trade flows to and from the Venetian marketplace, which stood at the junction between Western Europe and the Eastern Mediterranean (Mueller 1997). In order to finance their own commercial activities in Venice, companies like the Medici bank started to offer to Florentine investors this new type of financial product– *de facto*, a placement in Florentine currency whose return was indexed on Venetian interest rates (De Roover 1974). International trade to and from Venice was thus financed by Florentine capital, but only through the interface of Florentine companies and through the issuance of non-marketable local credit contracts.

## 2.2. The rise of negotiable trade finance instruments

The nature of the bill of exchange changed considerably in the early modern era with the introduction of negotiability, which transformed the bill from a mere certificate of a local credit contract into an exchange-traded financial instrument. This evolution followed from fundamental juridical changes that took place in the Low Countries in the early $16^{th}$ century and were spread throughout Europe thanks to the leading role of first Antwerp and then Amsterdam as international commercial hubs. These developments transformed the trade finance sector considerably. The introduction of negotiability fostered the decline of the medieval network companies and allowed for importing and exporting firms to lend directly to each other by purchasing each other's bills. Therefore, by the mid-$18^{th}$ century, the financing of commercial transactions mostly took place though credits granted *between firms* themselves specializing in overseas commerce, and the role of bankers in the intermediation of such credits had declined.

The crucial steps towards these transformations took place in Antwerp in the first half of the $16^{th}$ century (Braudel 1982). From the 1510s until the 1560s, Antwerp played the role of commercial metropolis of Western Europe. After supplanting Bruges' early role as a regional hub and breaking

Venice's monopoly on the spice trade, the harbour on the river Scheldt became the point of junction between several transcontinental flows (from England, the Baltic, Central Europe, the Mediterranean, and Iberia), and remained so for around half a century (Van der Wee 1963). At the Antwerp fairs, the new model of "sedentary trade" (pioneered by Italian network companies like Bardi or Medici and imitated by South German groups like Fugger or Welser) met with the old model of "caravan trade" (still practised by English and Hanse merchants). While network companies could rely on a sophisticated management of inter-group liquidity for the financing of their operations, more primitive traders needed to mobilize quickly the proceeds of their sales in order to be able to convert them as soon as possible into merchandise for re-export. To this aim, Northern merchants insisted on obtaining from the Antwerp authorities the recognition of the principle of negotiability of credit instruments, which transferred to the bearer the juridical protection previously granted exclusively to the original creditor. Charles V's decrees of 1537 sanctioned this principle in the Low Countries, and in the following decades the practice of endorsement was gradually extended to bills of exchange (De Roover 1953, pp. 94-100). Negotiability was the necessary condition for trade finance to escape its purely local dimension. Once the rights of the original creditor were allowed to be transferred entirely to the bearer, the latter actually became the new creditor: from a loan extended in the place *from* which the bill was drawn, the bill of exchange became a loan extended in the place *on* which the bill was drawn (Geva 2011, pp. 352-422).

Given the role of Antwerp as the new European commercial hub, the "Antwerp custom" gradually imposed itself as the new international standard: by the early $17^{th}$ century, the negotiable bill of exchange had become the staple instrument for financing intra-European trade, allowing for a complete disjunction between the location of the borrower and that of the lender. From a formal viewpoint, the principle of negotiability of the bill of exchange (which stood in contrast to the medieval juridical tradition) was adopted by national legislations much more slowly (De Roover 1953, pp. 100-118). The "Antwerp custom" continued to provide the benchmark for legislation long after the Flemish city had been replaced by Amsterdam (whose commercial laws were the same as Antwerp's) as Western Europe's commercial metropolis; in France, for instance, it was integrated almost literally into the Code Savary of 1673, which later provided the basis for the Napoleonic Code de Commerce.

In the meantime, the emergence of Amsterdam as the leading commercial hub contributed to transforming the bill on Amsterdam into an increasingly popular instrument (Gillard 2004). Contrary to the late medieval age (in which the list of banking places, determined by oligopolistic Italian companies, was very limited), during the 17$^{th}$ and 18$^{th}$ centuries the infrastructure allowing for the financing of trade became free-entry, both in terms of geography and demography (Flandreau *et al.* 2009).

However, while the introduction of negotiability changed the structure of the trade finance market and trade finance products evolved considerably in that period, their circulation still remained relatively limited. The new negotiable bill of exchange appears to have circulated mostly among merchant firms themselves for the payment of reciprocal debts. While bills of exchange could well be vehicles of financial contagion - as it was spectacularly the case during the crisis of 1763- contagion seemed to spread mostly to agents themselves specializing in international commodity trade (Schnabel and Shin 2004; Santarosa 2015; Quinn and Roberds, 2015). International commerce was still mostly financed on a decentralized basis, through credits extended by specialized agents in the exporters' and importers' countries.

## 3. Trade finance during the first globalization and the bill on London

From the second half of the 18$^{th}$ century to the early 20$^{th}$ century, the emergence of a large discount market for bills of exchange in London and its progressive internationalization led to a profound transformation of the global trade finance market. Bills of exchange drawn on the London City started being used to finance commercial transactions taking place across the whole globe and circulated as highly liquid and safe money market instruments. A great diversity of domestic and foreign investors now purchased sterling bills on the London market and therefore contributed to finance international commerce. The global trade finance market became increasingly concentrated around one financial centre, where highly-standardized products were issued by specialized agents but bought by all kinds of investors. These evolutions allowed for the transformation of trade finance from an "insiders'" sector dominated by specialized firms to a broadly-accessed one. They also gave Britain a central role in financing the global trade boom of the second half of the 19th century and in regulating firms' access to trade finance.

One important step in the rise of London as a trade finance centre was the creation of large discount market for bills of exchange during the second half of the 18th century (De Roover 1953, pp. 129-142). As the author of the reference 18th-century work on English business, Malachy Postlethwayt, pointed out in 1751, it was around a brand-new instrument – the so-called *inland bill*- that the London discount market initially emerged (De Roover 1953, pp. 139-140). The inland bill was a bill of exchange drawn and payable in England and was therefore a purely domestic credit instrument. Its emergence was made possible by changes in the English jurisdiction regarding credit contracts which extended the principle of negotiability to domestic credit certificates. This widespread adoption of negotiability in England did not come without difficulties. When in 1622 the English merchant Gerard Malynes wrote his famous treatise on "*lex mercatoria*" (Malynes 1686), one of his main aims was to prevent common law courts from making domestic commercial practices diverge from the new international ones that had developed in Antwerp. As a matter of fact, mercantile courts had been abolished in early 17th century England, and commercial cases were now ruled by central courts (Rogers 1995, pp. 20-26). At the time Malynes was writing, the latter were hindering local adoption of negotiability, thus preventing domestic merchants from mobilizing their credits as they could do on the Continent (De Roover 1953, pp. 109-113). Pace Malynes, legal standards regarding credit certificates developed on a parallel track in England and on the Continent, but the final outcome was not substantially different – except for some minute details (Geva 2011, pp. 423-466). While English courts were initially more reluctant than Continental lawmakers to recognize the principle of negotiability, they lately proved more aggressive than the latter in extending it to all forms of private credit certificates and, in particular, to purely domestic ones (Richards 1929, pp. 44-49).

The inland bill became a very popular means of payment for domestic transactions in England where it was exchanged on an increasingly large discount market. By the end of the 18th century, the discount market of London became the national money market of the burgeoning English economy and the place where demand and supply of short-term credit met in England. Country banks with surpluses kept funds with London correspondents, which were typically invested in bills supplied by country agents with deficits (Scammell 1968, pp. 115-130).

Although the discount market developed in England as an essentially domestic market, it was soon used for financing international trade as well. By the end of the Napoleonic Wars, however, London had supplanted Amsterdam not only as an intra-continental, but also as an inter-continental commercial hub. At that time, intercontinental trade was still financed differently from intra-European trade. While the negotiable bill of exchange prevailed for the financing of European trade, commercial transactions between continents (which obviously required immobilizing capital for much longer periods) were still financed through idiosyncratic arrangements as, for example, the long-dated bills issued by "agency houses" in the Anglo-Indian trade (Chapman 1984). During the first half of the 19$^{th}$ century, however, instruments used to finance domestic, intra-continental, and inter-continental trade were progressively homogenised, so that bills of exchange drawn by agents located anywhere in the world became traded on the London discount market. Already in the 1850s, the majority of the bills discounted on the London market were drawn by borrowers located outside Britain, and a large share of these "foreign bills" originated from outside Europe (Flandreau and Ugolini 2013). The London discount market progressively became an almost exclusively international market and the inland bill disappeared from circulation by the beginning of the 20$^{th}$ century (Nishimura 1971).

Figure 2 gives an example of how bills on London (or acceptances) could be used to finance a commercial transaction between two merchants located anywhere in the world. An exporter in country A had sold goods to an importer in country B and needed to finance production and/or shipment. The importer could instruct the exporter to draw a sterling bill (covering the amount of the sale) on a London acceptor with which she had a relationship. The acceptor "accepted" the bill by putting its signature on it (in exchange for a fee) and, in so doing, she committed to repay the bill holder at maturity. Once the bill carried the signature of a reputable acceptor, the exporter could discount it against cash on the London discount market at the current market interest rate. The bill could change hands several times in London and be purchased by various types of investors. Once arrived at maturity, its current holder asked for payment in sterling to the acceptor, who, in the meantime, had arranged to obtain payment from the importer. Bound to pay bills of exchange at maturity even if they did not receive payment from the importer, acceptors were the *de facto* guarantors of the loans extended by discounters to the exporters/drawers.

**Figure 2. The bill on a London acceptor (or acceptance)**

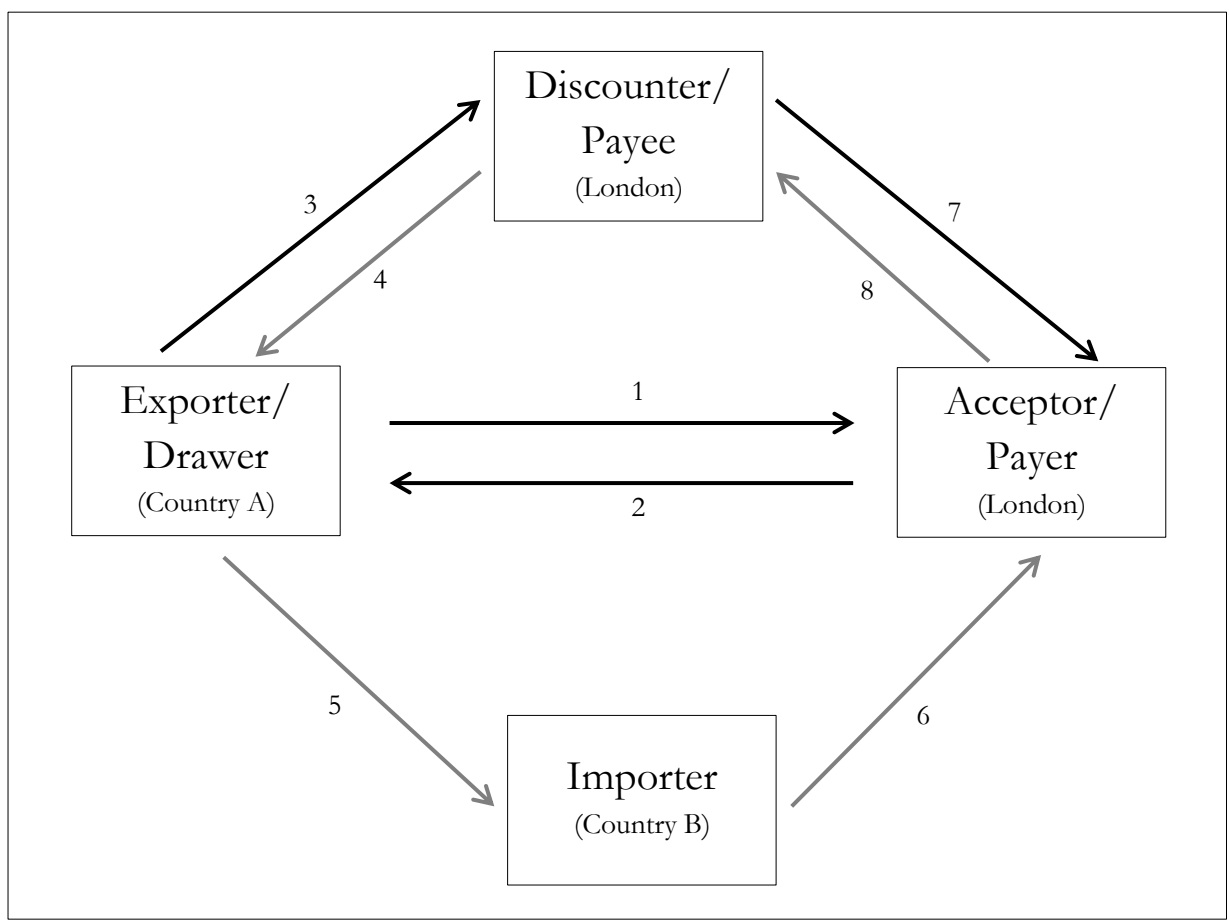

1. Draws a bill; 2. Accepts the bill; 3. Sells the accepted bill; 4. Discounts the bill/provides cash; 5. Ships goods; 6. Makes payment ("provision") before maturity; 7. Presents the bill at maturity; 8. Pays the bill at maturity/provides cash.

Standardization of credit instruments negotiated on the London discount market was made possible by the emergence of an informal set of governance rules, enforced through the attractiveness of this credit platform and the power of a few actors involved in regulating outsiders' access. In particular, specialized intermediaries known as "*acceptance houses*" emerged in London at that time. These institutions, whose founders often originated from the continent, had close ties with foreign countries and guaranteed (accepted) bills on account of their customers abroad in order to allow them to borrow from investors in the London bill market. Acceptance houses were not the only institutions accepting bills of exchange in that period as bills drawn on merchant firms continued to circulate and British commercial banks also provided acceptance facilities to their customers. However, the acceptance houses' connections abroad and extensive knowledge of merchants and firms in foreign countries allowed them to accept bills on account of a large number of foreign customers. By screening potential drawers/borrowers and establishing which of them were eligible for their guarantee, they acted as "gatekeepers" to the London discount market and regulated firms' access to the most important source of trade finance globally (Accominotti, Lucena, and Ugolini forthcoming).

Investment in the bill market was also encouraged by the Bank of England's adoption of lending-of-last-resort policies through its discount window, which guaranteed the ultimate liquidity of eligible bills under any market conditions. By setting formal and informal eligibility rules to discount (and also by monitoring the discounters), the Bank of England encouraged the orderly production of credit instruments on the market. Therefore, the British central bank used its market power (i.e., its monopoly of emergency lending in crisis time) as a means for fostering standardization on the London bill market. In so doing, the Bank actually contributed to decreasing the credit and liquidity risk associated with bills, and thus supported the role of the bill on London as an international "safe asset" (Flandreau and Ugolini 2014).

In the period from 1870 to 1914, London was the global centre for the financing of international trade and its discount market morphed into *the* global money market. During those years, the role of London as the world's commercial hub remained largely unrivalled, and the adoption of the international gold standard made the bill on London – drawn from any country, but payable by a London acceptor and eligible for rediscount at the Bank of England – the most widely demanded short-term financial instrument in the world (Lindert 1969; Flandreau and Jobst 2005). A great variety of investors contributed to finance the global trade boom of 1870-1914 through purchasing sterling bills. The increasing depth of the London discount market at the beginning of the 19th century encouraged the emergence of the "bill brokers" or discount houses, who specialized in buying bills of exchange and financed themselves through short-term deposits and credits from commercial banks (King 1936). Other English financial institutions and investment trusts, as well as foreign central and commercial banks, also invested in sterling bills. This evolution involved a disconnection between the location of importing and exporting firms and that of the lenders. Through the platform of the London market, British and foreign investors could now lend funds to borrowers located in any country and finance a commercial transaction occurring in any part of the globe – not necessarily involving a trade flow to or from London.

The London market gave firms wide access to trade finance services but entry was not completely free. Local "gatekeepers" and "monitors" had the power to regulate access to this central platform and extracted profits from their position. In particular, the London acceptance houses benefitted from the fees they charged in exchange for their signature/acceptance of bills. The London discount market was therefore at the centre of the system of British Global Governance that emerged before the First World War.[2] The position of Britain in the global trade finance market provoked the admiration and jealousy of foreigners and aroused the envy of potential competitors who, towards the end of the period, started to figure out how to end London's monopoly and the "tribute" paid to it (Broz 1997).

It is remarkable that the substantial transformation experienced by the London discount market between the mid-18th century (when it emerged as a purely domestic platform) and the late 19th century (when it became a purely international one) took place without any formal modification in the legal status of the bill of exchange: the one important juridical initiative of the period (the Bills of Exchange Act of 1882) only limited itself to ratifying the already existing practices and rules codified by the common law courts during the 17th and early 18th centuries (Geva 2011, pp. 528-584). Although their importance in financing international trade remained limited, other discount markets modelled after the English one also developed on the European Continent in the course of the 19th century. For these markets to emerge, legal restrictions to the circulation of purely domestic credit instruments, previously existing in many places (e.g. Venice or Amsterdam), had had to be lifted with the end of the Ancien Régime (Ugolini 2017, pp. 55-61). The legal benchmark for bills of exchange was, however, provided there by the Napoleonic Code de Commerce, which still faithfully followed the original "Antwerp custom". Thus, one might legitimately say that the only significant legal change that took place in the long history of Western trade finance was actually the introduction of the principle of negotiability in 16th-century Antwerp. One should not, however, underemphasize the role of informal regulation setting eligibility standards for access to the market by outsiders. Both acceptance houses and the Bank of England contributed to set such standards, and informal regulation was binding proportionally to its proponents' market power. This was also the case on the Continent, where merchant banks and banks of issue established domestic standards by modifying eligibility criteria (Ugolini 2017, pp. 134-143).

---

[2] Britain's central position in the regulation of global trade finance echoes its importance in setting standards and regulating international commodity trade itself. See Sgard (2018) for a discussion based on the case of the London corn market in 1885-1930.

## 4. The disintegration of global trade finance, 1914-1939

### 4.1. The First World War and disruptions in the London money market

The role of London in the global trade finance market progressively declined during the First World War (WW1) and interwar years. The war itself resulted in severe disruptions in the functioning of the London discount market. John Maynard Keynes (1914) described in September how the political developments of the summer 1914 affected the City. Even before Britain declared war on Germany in August 1914, the July crisis led several continental countries to declare moratoria on foreign exchange payments and to close their Stock Exchanges. The immediate consequence of these capital controls was that foreign debtors who had drawn bills of exchange on the leading acceptance houses or banks of London could not remit funds to these institutions in order to reimburse their credits (Keynes, 1914; Roberts, 2013). The London acceptance houses were however bound to repay the holders of these bills (the ultimate lenders) in pounds sterling at maturity. In effect, this meant that they had to draw on their own liquid assets in order to honour the guarantees on trade finance credits granted to foreign customers. The outcome was disastrous for houses that had accepted/guaranteed large amounts of bills relative to their liquid or total assets and several of them were simply unable to assume their liabilities. The possibility of their failure could have resulted in a complete breakdown of the machinery of the London discount market.

In order to avoid a collapse of the British financial system, the British Government and Bank of England reacted strongly (Roberts, 2013). Emergency measures were soon passed to allow British acceptance houses first to postpone payment of all bills coming due, and second, to borrow directly from the Government for a period that could be prolonged until the end of the hostilities (Greengrass, 1930, p. 102; Spalding, 1933, pp. 241-245). These measures avoided Britain a banking crisis. However, for the first time, the liquidity of the world's most widespread trade finance instrument – the sterling bill – had been put into question. In addition, restrictions on capital flows and the difficulties of conducting international commerce in times of war severely weakened the role of London as a global trade finance centre.

**4.2. Revival of trade finance in the 1920s and dual market structure**

Following WW1, the global trade finance market was gradually reconstructed on the basis of the same instruments that had prevailed in the nineteenth century. After the stabilization of most European currencies in the mid-1920s, the second half of the decade saw a revival in international trade and capital flows and the demand for credit from importing and exporting firms increased considerably. However, London's hegemony was challenged in those years by the emergence of another large acceptance market in New York. The centralised structure of the global trade finance market that had prevailed before WW1 therefore gave way to a dual structure where London and New York competed actively for the financing of international commerce.

The rise of the New York centre was a natural consequence of WW1 from which the United States emerged as the largest creditor country and the world's leading commercial power. The origins of the New York acceptance market dated back to the years 1908-1912 when, following the 1907 banking panic, US authorities mandated a special commission (the National Monetary Commission) to produce reports on various foreign countries' banking systems. The unassailable predominance of London in international trade finance in the nineteenth century provided one of the main motivations for financial reforms which eventually led to the foundation of the Federal Reserve (Warburg 1910). A well-known banker unofficially advising the Commission, Paul Warburg, argued that the absence of a large money market in the United States similar to the London one resulted in severe inefficiencies (Warburg, 1914). Restrictions on US national banks' foreign banking and acceptance activities before WW1 and the lack of a liquid market for trade finance products obliged American firms to finance their international trade transactions via the London market. This involved considerable costs for those firms and gave British banks the possibility to regulate American firms' access to financing facilities.

As of 1913, US monetary authorities therefore attempted to mimic the main features of the London money market (Ferderer, 2003; Eichengreen, 2010; Eichengreen and Flandreau, 2012). The Federal Reserve Act of 1913 removed restrictions on US national banks' and Federal Reserve member banks' acceptance activities. In the following years, a market for dollar denominated bankers' acceptances developed in New York (Ferderer, 2003). In contrast to London where the largest share of bankers' acceptances was issued by small acceptance houses specialized in trade finance, in the United States, the business of accepting bills of exchange was dominated by the country's largest commercial banks (Accominotti, 2018). The US acceptance market also received direct support from monetary authorities as the Federal Reserve Banks bought a very substantial share of acceptances issued by US banks in the 1920s (Eichengreen and Flandreau, 2012). At the end of the 1920s, the volume of dollar bankers' acceptances increased considerably and New York started competing with London for the financing of world trade.

London, however, had not said its last word. When continental European commerce resumed following the currency stabilizations of the 1920s, a large share of the associated credit demand still remained directed to the City (Greengrass, 1930, p. 30). London's geographical location, the structure of its money market, and the expertise of the British acceptance houses in intermediating credit for foreign merchants and firms remained clear advantages in the trade finance business.

**Figure 3. Share of world trade financed by US and UK bankers' acceptances, 1927-1935 and 1951-2000**

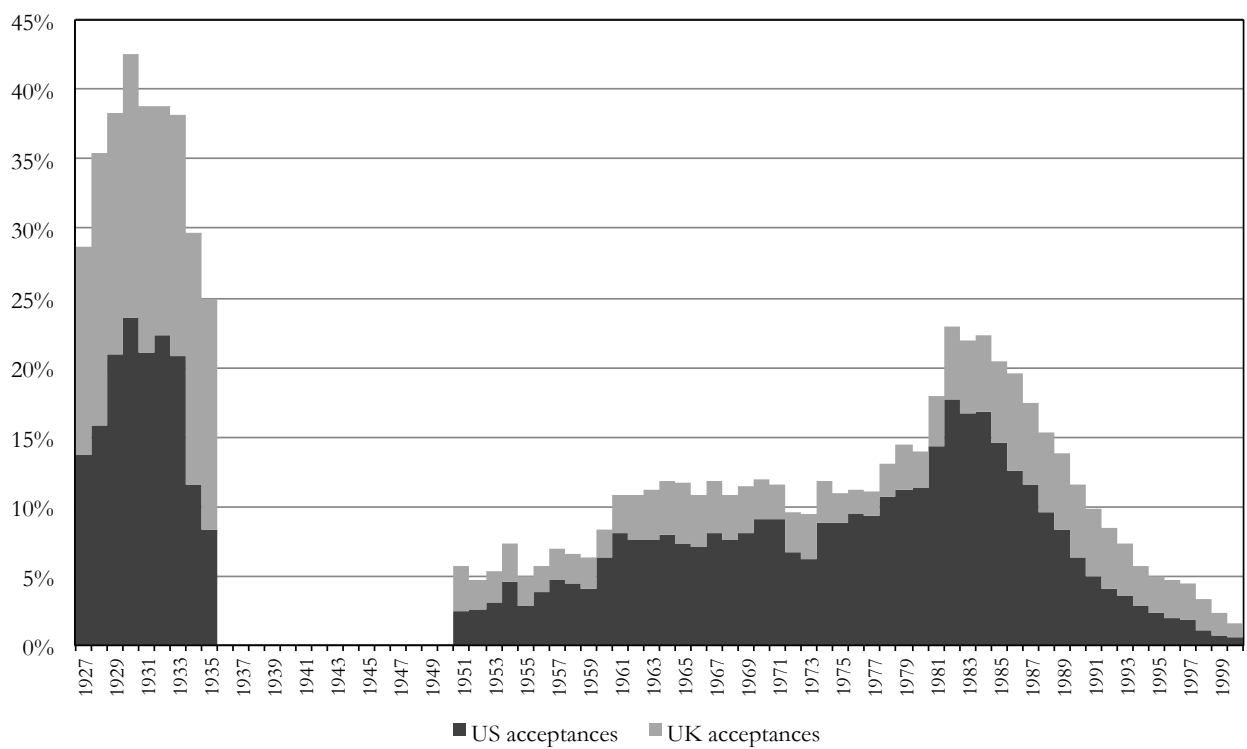

Sources: UK acceptance outstanding are from Baster (1937) for 1927-1935, the Bank of England's *Quarterly Bulletin* (various issues) for 1950-1978 and the Bank of England's Statistical Interactive Database for 1979-2000. US acceptances outstanding are from Carter et al. (2006), series Cj1185 for 1927-1996 and from the Board of Governors of the Federal Reserve System's *Statistical Digest, 1996-2000* for 1997-2000. World exports are from Federico and Tena (2016) for 1927-1935 and from the International Monetary Fund's *International Financial Statistics* for 1950-2000. An average maturity of 90 days is assumed for bankers' acceptances.

Figure 3 displays the share of world trade financed through acceptances drawn on US and UK banks, annually, from 1927 to 1935 and from 1951 to 2000. The figure clearly illustrates the dual structure of the global trade finance market in the 1920s. In 1930, New York and London were financing equal shares of global trade; as much as 43 per cent of world exports were financed through acceptances drawn on banks in these two financial centres. This intense competition between London and New York banks in the global acceptance market led to a lowering of the fees they charged in exchange for their guarantees and contemporaries complained that it resulted in a cutback in the standards set by intermediaries and monetary authorities to regulate access to trade finance services (Baster, 1937, p. 300).

**4.3. The collapse of trade finance in the 1930s**

The revival of acceptance financing of the 1920s was however short-lived. The world economic crisis of the 1930s soon contributed to a disintegration of the existing structures of international trade finance. One major driving force behind this development was the collapse of world trade that occurred during the Great Depression due to the decline in world income and increase in trade tariffs (Kindleberger, 1976). Between 1929 and 1933, world exports declined by almost 30 per cent in real terms and this considerably reduced the *demand* for financing from firms (Federico and Tena, 2016).

At the same time, the financial crisis of the 1930s also affected the *supply* of trade finance. A wave of financial crises and the imposition of capital controls in Germany and Central European countries in the summer of 1931 jeopardized the reimbursement of bills drawn on London and New York banks by firms located in the region and resulted in the freeze of the US and British banks' Central European credits (Harris, 1935; Ellis, 1941). Once again, the acceptance houses of the London City were the most severely affected by the international economic situation because they strongly specialized in intermediating trade finance for continental customers. These houses experienced a run on their deposits and their balance sheets contracted (Accominotti, 2012). In the United States, by contrast, the acceptance business was mostly dominated by the largest commercial banks, which were more diversified and therefore less impacted by the European crisis (Accominotti, 2018). However, perhaps in an effort to strengthen the credibility of the dollar's gold parity, the Federal Reserve apparently withdrew its support to the New York acceptance market in 1931 and considerably reduced its holdings of bankers' acceptances. This put the expansion of the New York market to a hold (Eichengreen and Flandreau, 2012).

The emergence of regional trading blocks in the second half of the 1930s resulted in a complete reorganization of the global trading system. Many countries imposed quantitative restrictions on trade flows and bilateral clearing agreements with Central European and Latin American countries proliferated. These agreements were based on the principle of reciprocal trade and left the management of the bilateral trade balance to a government agency or compensation office (League of Nations, 1935). They therefore resulted in increased state interference in international commerce. In addition, capital controls as imposed by many countries' governments in the 1930s reduced the scope for extending cross-border credits and led to a further disintegration of the global trade finance market. In 1935, the New York and London discount markets only financed 25 per cent of world exports (as opposed to 43 per cent in 1930).

At the same time, trade finance products had ceased to be the keystone of the London money market. During WW1, British authorities had issued an increasing volume of Treasury bills. The Treasury bill - backed by government debt - replaced the bankers' acceptance - backed by private firms' debt - as the most important money market instrument in the interwar period (Greengrass, 1930; Truptil, 1936).

Ironically, just a few months before the start of the global economic crisis that would result in the weakening of the bill of exchange system, this very system received its first formal juridical codification at the supranational level. In 1930, the Convention on the Uniform Law of International Bills of Exchange was signed in Geneva. This initiative was promoted by the League of Nations and was one of the very first attempts at establishing a uniform international financial regulation.[3] The Convention was essentially based on the Continental legal tradition. It was relatively similar to the English convention but formally differed from it in a number of details. In view of these formal differences, however, Anglo-Saxon countries refused to ratify the Convention, thus preventing the establishment of a uniform international legal standard. This legal loophole in the regulation of payment instruments persists to date.[4]

---

[3] To be precise, the process had been started before the WW1 with the two conferences of The Hague in 1910 and 1912, but it had been put on hold by the outbreak of the conflict (Moshenskyi 2008, pp. 170-172).

[4] In 1988, the United Nations Commission on International Trade Law (UNCITRAL) tried to fill this loophole by seeking convergence between the Anglo-American and Continental legal traditions. Although approved by the General Assembly on 9th December 1989, the resulting United Nations Convention on International Bills of Exchange and International Promissory Notes never entered into force, having been ratified by no country to date (Murray 1994; Moshenskyi 2008, pp. 172-174).

# 5. The reconstruction of trade finance after WW2

## 5.1. The era of state-regulated trade and finance

The outbreak of the Second World War (WW2) finalised the disintegration of the world's trading system and trade finance market that had been initiated during the 1930s. The war was accompanied by further restrictions on international commercial and financial transactions. When international trade resumed at the end of WW2, the old channels to finance commercial activities were reactivated. The London merchant and clearing banks as well as the main American commercial banks resumed their business of granting trade credits to exporters and importers - both at home and abroad- through acceptances.

In doing so however, suppliers of trade finance were confronted with new challenges. Under the newly established Bretton Woods system, international payments were subject to tight government regulations. The Bretton Woods conference of 1944 organized a system of fixed exchange rates where the US dollar was the dominant currency, and countries maintained restrictions on international capital movements (Eichengreen, 1996). Many of the bilateral clearing agreements that had been put into place in the 1930s were also maintained alive after the war. These combined measures reinforced the principle of a state regulation of international payments and trade.

Institutions engaging in trade finance had to deal with these constraints and adapt their activities to the new regulatory environment of the post-war years. In the United Kingdom, any credit granted to a foreign resident or British company in foreign ownership in the early post-war years had to be authorized by the exchange control (Bank of England, 1967, p. 250; Steffenburg, 1949, p. 74). Financial institutions engaging in trade finance devoted considerable efforts to ensure that the international transactions they financed were done in accordance with government regulations. For example, they had to verify that every importer seeking a short-term credit had been issued a valid import license. E. Steffenburg, a partner at the London merchant bank Hambros, noted in 1948 how the compliance with domestic and foreign regulations was the most time-consuming task performed by his bank's staff in the post-war years (Steffenburg, 1949, p. 66). Although restrictions to foreign lending were progressively removed over the 1950s and 1960s and current account convertibility was restored, international trade and finance remained regulated by state authorities until the end of Bretton Woods.

**5.2. The global trade finance market in the post-Bretton Woods years, 1973-1985**

Following the collapse of the Bretton Woods system in 1971-1973, capital controls were progressively removed and a revival in international trade in the late 1970s and early 1980s resulted in increased demand for trade finance. Since the United States was now at the centre of the global trading system and the US dollar had consolidated its status as the dominant international currency, it is mostly through the US market that exporters and importers of the whole globe sought to finance their commercial activities in the post-Bretton Woods years. The US bankers' acceptance market experienced a new boom in the late 1970s. This development was facilitated by US banking regulation. Acceptances were eligible for discount as well as open market purchase by the Federal Reserve Board and they could also be used as collateral for advances. In addition, they became exempt from reserve requirements in 1973, making the activity of accepting bills an attractive business for American banks (Jensen and Parkinson, 1986; LaRoche, 1993).

US bankers' acceptances accounted for a substantial share of the financing of global trade in the 1970s and 1980s (Hervey, 1983; Jensen and Parkinson, 1986) and, in 1982-1984, around 17 per cent of world exports were financed through this instrument (figure 3). Despite this revival, however, the New York market never regained the importance it had had at the end of the 1920s in the financing of world trade. Starting from 1985, the issuance of US dollar bankers' acceptances steadily declined and, at the beginning of the 2000s, they only financed a negligible share of global exports (figure 3). Several reasons explain this decline. First, with the deregulation of financial markets, substitutes for acceptances developed for financing merchandise trade in the 1980s. In particular, the growth of the commercial paper market allowed large corporations to borrow directly from non-financial investors and without the signature/guarantee of a US money-center bank (Jenson and Parkinson, 1986, p. 10). In addition, acceptances lost their privileged regulatory status in the United States when the Federal Reserve Board of Governors also exempted other forms of short-term assets such as asset-backed commercial paper from capital requirements at the end of 1990 (LaRoche, 1993, p. 82). This made the issuance of acceptances a less attractive activity for US financial institutions. American banks therefore played an increasingly marginal role in the financing of world trade from the 1980s onwards.

**5.3. International trade finance in the first and second globalization**

The failed revival of the acceptance market in the post-war period left the second globalization of the late 20$^{th}$ century with a trade finance infrastructure that is profoundly different from that which prevailed during the global trade boom of the 19$^{th}$ century. While the methods of financing commerce have evolved profoundly, trade finance nowadays appears to be mostly intermediated by local banks in the exporting and importing firms' countries - as it was the case at the very origins of this market- rather than through a global platform such as that developed in London over the course of the 19$^{th}$ century.

Exporters and importers around the world nowadays finance their activities in various ways. They first rely on inter-firm trade credit, in which case the commercial transaction is financed directly by the exporter (*open account* method) or pre-paid by the importer (*cash in advance*/pre-payment method). However, national and global banks also offer various forms of trade finance services. They provide direct loans and overdraft facilities to firms in need of working capital (Cooper and Nyborg, 1997; Amiti and Weinstein, 2011) and large corporations also borrow in the US or Euro commercial paper markets (Asmundson et al., 2010). Finally, banks offer specific trade finance products aiming at insuring exporters against their importers' default risk, the most common of which are the *letter of credit* and *documentary collections* (Amiti and Weinstein, 2011; Bank for International Settlements, 2014; Niepman and Schmidt-Eisenlohr, 2016, 2017).

**Figure 4. The confirmed letter of credit**

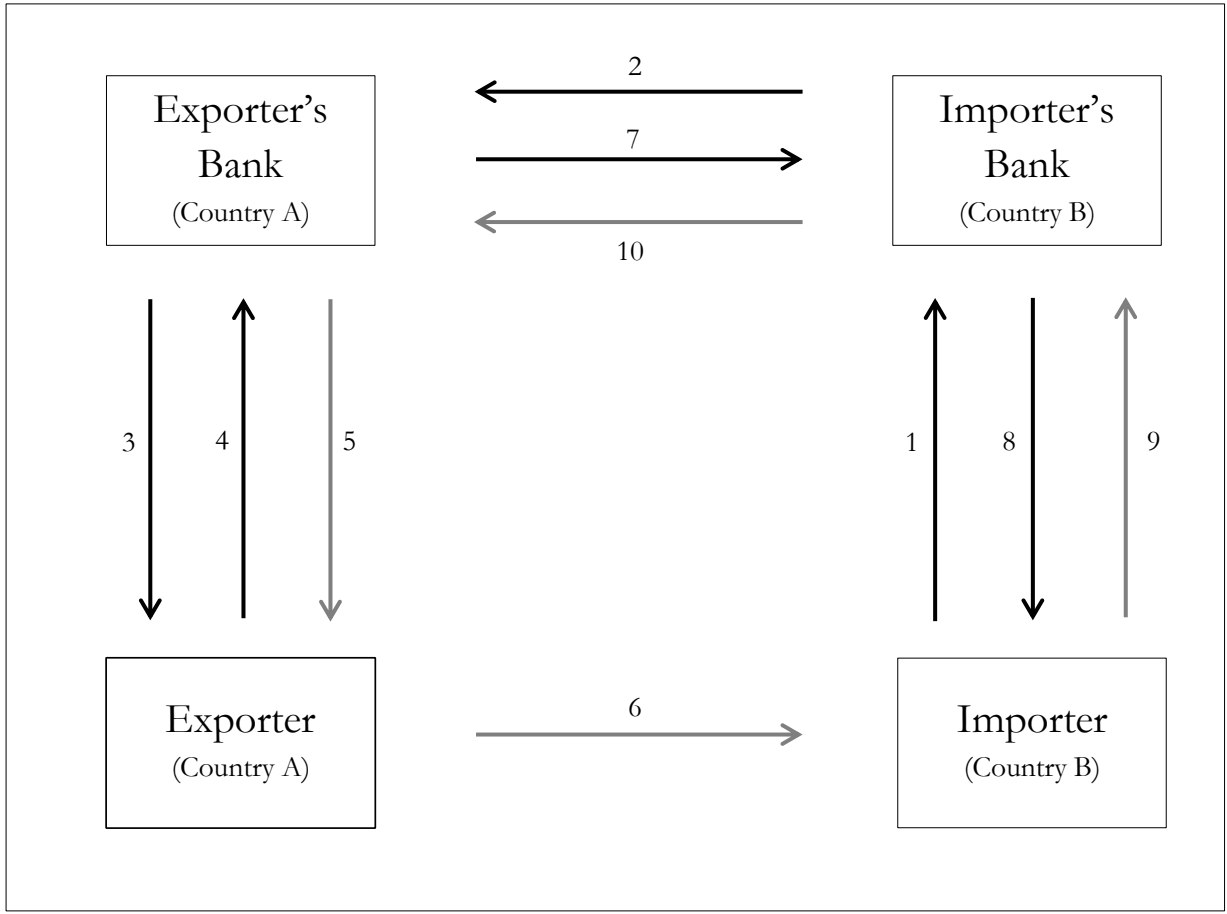

1. Applies for letter of credit; 2. Sends letter of credit; 3. Confirms letter of credit; 4. Sells letter of credit; 5. Discounts letter of credit/provides cash; 6. Ships goods; 7. Sends documents; 8. Presents documents at maturity; 9. Pays at maturity/provides cash; 10. Credits at maturity.

A letter of credit (or documentary credits) is an instrument, which allows a bank to guarantee an importing firm's payment to an exporter. The importer's bank issues a letter of credit guaranteeing the exporter that payment will be made upon presentation of a set of documents providing evidence that the merchandise has been shipped. The letter of credit is often "confirmed", in which case payment is also guaranteed by the exporter's bank. While a letter of credit in principle only consists in a payment guarantee granted by the exporter's and importer's banks, it is also often coupled with a working capital loan. Upon acceptance of the letter of credit by the issuing bank, the exporter often sells it to her own bank at a discount, therefore obtaining cash before receiving payment from the importer (Amiti and Weinstein, 2011, pp. 1845-1846 and figure 4). The operation in that case is substantially (albeit not formally) similar to that involved in a medieval bill of exchange (figure 1): the exporter borrows from her local bank upon presentation of a claim on the importer, and the lending bank recovers its credit through its correspondence with the importer's bank abroad. In the case of documentary collections, by contrast, banks do not guarantee payment nor lend working capital but only provide assistance to the exporter and importer in completing the transaction through handling documents and assisting in the collection of payment (Asmundson et al., 2010; Bank for International Settlements, 2014; Niepman and Schmidt-Eisenlohr, 2017). Finally, exporters can purchase insurance from non-bank, insurance companies against their importer's payment default or obtain guarantees from public export credit agencies (Bank for International Settlements, 2014).

The Bank for International Settlements (2014, p. 8) estimates that around one third of world trade was financed through trade finance products offered by banks in 2011, while the rest was financed directly by exporting and importing firms.[5] Niepman and Schmidt-Eisenlohr (2017) estimate that approximately 13 per cent of global trade is financed through letters of credit and an additional 1.8 per cent through documentary collections. They also show that letters of credit are used more frequently when the value of the payment guarantee is high, for example in transactions consisting in an import of goods to a low-income country.

The structure of the twenty-first century's global trade finance market differs from that of the first globalization in several dimensions. First, trade finance nowadays is mostly intermediated on a local basis. While in the nineteenth century commercial transactions taking place in any part of the globe were financed through the London financial centre, trade finance products are nowadays mostly intermediated through national banks or branches of global banks located in the exporter's and importer's country. Local or regional banks accounted for the majority of the global trade finance supply in 2011 (Bank for International Settlements, 2014, p. 11). As a result, the regulation of the global trade finance market is much more decentralized nowadays than it used to be in the first globalization. While the London acceptance houses – in connection with the Bank of England – defined the standards governing firms' access to trade finance in the nineteenth century, such standards are now set by banks at the local level. Centralization around London before WW1 involved that most banks offering trade finance services were under the implicit regulation of the Bank of England, whereas regulation of local and global banks is now left to national authorities. This difference in the structure of the trade finance market mostly reflects the more decentralized structure of the global trading system in the second than in the first globalization. The decentralized organisation of international trade finance nowadays somehow resembles that which prevailed at the origins of this market in the medieval period. However, even in periods when trade finance was mostly intermediated locally, global actors have been involved in its provision as they maintained a network of branches in local markets. Between one-fourth and one-third of trade finance today is provided by the local branches of global banks (Bank of International Settlements 2014, p. 11). This situation echoes the international pre-eminence of Italian and South-German banking groups in the late Middle Ages (De Roover 1953).

---

[5] These estimates vary depending on the definition of "trade finance" adopted. For example, estimates based on bank surveys typically conclude to a higher share of global trade financed through banks because institutions surveyed usually report their exposure to working capital loans to exporters and importers and not only exposure to specific trade finance products. See Bank for International Settlements (2014, pp. 11-12) and Niepman and Schmidt-Eisenlohr (2017).

One implication of the recent market structure is that firms located in countries where the banking system is underdeveloped might suffer from a lack of intermediation, a phenomenon known as the "trade finance gap" (Asmundson et al., 2011). Firms' access to trade finance might also be more sensitive to shocks to the domestic financial system. For example, Amiti and Weinstein (2011) provide evidence that Japanese firms whose local bank endured more severe losses during the crisis of the 1990s reduced their exports more than others. At the same time, access to trade finance remains reduced during episodes of global financial crises if exporters finance themselves through branches of global banks or through domestic banks depending on foreign capital for their refinancing. In particular, many researchers have noted how the global financial crisis of 2008 affected the supply of trade finance and resulted in credit constraints for firms engaging in international commerce (Ahn, Amiti and Weinstein, 2011; Del Prete and Federico, 2014; Paravisini et al., 2015; and Niepman and Schmidt-Eisenlohr, 2016). This drying up of financing facilities partly contributed to the collapse in world trade that took place in the year following the crisis.

Another aspect of today's trade finance market is that credits are rarely securitized. The introduction of negotiability in the sixteenth century allowed bills of exchange financing commercial transactions to circulate much more widely than they had done previously. In the nineteenth century and interwar period - and to a lesser extent, in the 1970s and 1980s- bankers' acceptances drawn by firms around the world on leading financial houses in London and New York were used as money market instruments and purchased by a great variety of bank and non-bank investors. However, starting in the 1980s, the decreased use of trade finance products for money market transactions substantially narrowed the range of investors involved in the provision of trade finance. While global banks have recently attempted to securitize their trade finance portfolios or specific trade finance credits, demand for such products from non-bank investors remains limited due to the lack of standardization and knowledge of this type of products (Bank for International Settlements, 2014, pp. 27-30). The ultimate source of funding for international commercial transactions nowadays comes from either the trading firms themselves (in the case of inter-firm credit or when a letter of credit is issued but not discounted by the exporter's bank) or from their banks (in the case of a letter of credit coupled with a working capital loan). In contrast to the pre-WW1 period therefore, when investors in bills of exchange were of various types and did not necessarily have any specific knowledge of the ultimate borrowers, providers of capital for the financing of international trade nowadays mostly consist of local banks who are in direct contact with borrowing firms.

## 6. Conclusion

This chapter has presented a long-term perspective on international trade finance with a special focus on the structure and global governance of this market. We found that similar trade finance instruments have remained in use over more than eight centuries. The bill of exchange, which traces its roots back to the Antiquity but only became common use in the West in the 13th century, has survived under different forms in the global economy and is still used today for financing commercial transactions. This instrument has proven to be extremely flexible and the variety of its uses corresponded to major evolutions in the structure of the global trade finance market. Medieval bills of exchange were idiosyncratic instruments designed by a few local merchant bankers corresponding between different cities. They were therefore instruments of localised bank-intermediated trade finance. During the early modern period, the appearance of the negotiable bill of exchange allowed trade finance instruments to circulate more widely, although their distribution remained mostly limited to firms specializing in international commerce. The disconnection between the location of the borrower and that of the lender became complete during the 19th century, when the standardization of bills (or acceptances) turned the London acceptance market into the world's staple money market. With the disintegration of the global economy in the interwar and post-WW2 period, however, the acceptance market almost disappeared and failed to be revived after 1980 when international trade experienced a new boom. Whereas the global trade finance market was highly centralized around the City of London during the first globalisation, trade finance nowadays mostly consists in a range of products offered to firms by local banks or the local branches of global banks – as it was the case at the origins of this market.

This evolution in the structure of international trade finance has implications for the global governance of this market. In the nineteenth century, governance and regulation of trade finance was exercised by the leading political power of the time: Britain. Firms willing to access the facilities offered by the London discount market needed to obtain a signature from a reputable London financial institution and the British acceptance houses and commercial banks set the standards required to obtain such credits. At the same time, the main trade finance suppliers were located in London and subject to the (informal) regulatory authority of the Bank of England. The pre-eminence of Britain in the governance of the global trade finance market was however challenged during the interwar period when London acceptance houses had to compete with the large US commercial banks in the provision of trade finance services. Nowadays, criteria governing firms' access to trade finance credits are defined on a local basis, while regulation of banks offering trade finance services is left to national authorities. Whereas the international financial system that emerged in the nineteenth century allowed Britain to regulate trade finance globally, the more decentralized structure that prevails nowadays makes international control over the trade finance market less feasible and pushes back its governance into a sort of anarchy.

# References

Accominotti, Olivier (2012), "London Merchant Banks, the Central European Panic and the Sterling Crisis of 1931", *The Journal of Economic History*, vol. 72, pp. 1-43.

Accominotti, Olivier (2018), "International Banking and Transmission of the 1931 Financial Crisis", *The Economic History Review*, forthcoming.

Accominotti, Olivier, Delio Lucena, and Stefano Ugolini (2018), "The Origination and Distribution of Money Market Instruments: Sterling Bills of Exchange During the First Globalization", mimeo, LSE/Sciences Po Toulouse.

Ahn, JaeBin, Mary Amiti, and David E. Weinstein (2011), "Trade Finance and the Great Trade Collapse", *American Economic Review Papers and Proceedings*, vol. 101, pp. 298-302.

Amiti, Mary, and David E. Weinstein (2011), "Exports and Financial Shocks", *Quarterly Journal of Economics*, vol. 126, pp. 1841-1877.

Asmundson, Irena, Thomas William Dorsey, Armine Khachatryan, Ioana Niculcea, and Mika Saito (2011), "Trade and trade finance in the 2008-09 financial crisis", IMF Working Paper 11/16.

Bank of England (1976), "The UK Exchange Control: A Short History", *Bank of England Quarterly Bulletin*, Q3, pp. 245-260.

Bank for International Settlements (2014), "Trade Finance: Developments and Issues", CGFS Papers No 50, January 2014.

Baker, John H. (1979), "The Law Merchant and the Common Law before 1700", *Cambridge Law Journal*, vol. 38, pp. 295-322.

Baster, Albert S. J. (1937), "The International Acceptance Market", *American Economic Review*, vol.27, pp. 294-304.

Bogaert, Raymond (1968), *Banques et banquiers dans les cités grecques*, Leyden: Sijthoff.

Braudel, Fernand (1982), *Civilization and Capitalism, 15th-18th Century: The Perspective of the World*, Berkeley and Los Angeles: University of California Press.

Broz, J. Lawrence (1997), *The International Origins of the Federal Reserve System*, Ithaca NY and London: Cornell University Press.

Cerutti, Simona (2003), *Giustizia sommaria: Pratiche e ideali di giustizia in una società di Ancien Régime (Torino XVIII secolo)*, Milan: Feltrinelli.

Chapman, Stanley D. (1984), *The Rise of Merchant Banking*, London: Allen & Unwin.

Cohen, Edward E. (1992), *Athenian Economy and Society: A Banking Perspective*, Princeton NJ: Princeton University Press.

Cooper, Stuart, and Inke Nyborg (1997), "The Financing and Information Needs of Smaller Exporters", *Bank of England Quarterly Bulletin*, Q2, pp. 166-172.

De Roover, Raymond (1953), *L'évolution de la lettre de change, XIVe-XVIIIe siècles*, Paris: Armand Colin.

De Roover, Raymond (1974), "*Cambium ad Venetias*: Contribution to the History of Foreign Exchange", in Julius Kirshner (ed.), *Business, Banking, and Economic Thought in Late Medieval and Early Modern Europe: Selected Studies of Raymond De Roover*, Chicago and London: University of Chicago Press, pp. 239-259.

Del Prete, Silvia, and Stefano Federico (2014), "Trade and Finance: Is There More than Just "Trade Finance"? Evidence from Matched Bank-Firm Data", Bank of Italy Working Papers No 948, January 2014.

Ellis, Howard (1941), *Exchange Control in Central Europe*, Cambridge MA: Harvard University Press.

Eichengreen, Barry (1996), *Globalizing Capital: A History of the International Monetary System*, Princeton NJ: Princeton University Press.

Eichengreen, Barry (2010), *Exorbitant Privilege: The Rise and Fall of the Dollar and the Future of the International Monetary System*, Oxford: Oxford University Press.

Eichengreen, Barry, and Marc Flandreau (2012), "The Federal Reserve, the Bank of England, and the Rise of the Dollar as an International Currency, 1914–1939", *Open Economies Review*, vol. 23, pp 57-87.

Federico, Giovanni, and Antonio Tena-Junguito (2016), "World Trade, 1800-1938: A New Data-Set", EHES Working Paper No 93 (January 2016).

Ferderer, J. Peter (2003), "Institutional Innovation and the Creation of Liquid Financial Markets: The Case of Bankers' Acceptances, 1914-1934", *The Journal of Economic History*, vol. 63, pp. 666-694.

Flandreau, Marc, and Stefano Ugolini (2013), "Where It All Began: Lending of Last Resort and Bank of England Monitoring During the Overend-Gurney Panic of 1866", in Michael D. Bordo and William Roberds, *The Origins, History, and Future of the Federal Reserve: A Return to Jekyll Island*, New York: Cambridge University Press, pp.113-161.

Flandreau, Marc, and Clemens Jobst (2005), "The Ties That Divide: A Network Analysis of the International Monetary System, 1890-1910", *The Journal of Economic History*, vol. 65, pp. 977-1007.

Geva, Benjamin (2011), *The Payment Order of Antiquity and the Middle Ages: A Legal History*, Oxford and Portland OR: Hart Publishing.

Gillard, Lucien (2004), *La banque d'Amsterdam et le florin européen au temps de la République néerlandaise (1610-1820)*, Paris: EHESS.

Greengrass, H. W. (1930), *The Discount Market in London: Its Organization and Recent Developments*, London: Sir I. Pitman and Sons Ltd.

Hervey, Jack. L. (1983) "Bankers' Acceptances Revisited", Federal Reserve Bank of Chicago *Economic Perspectives*, vol. 7 (May 1983), pp. 21-37.

Jensen, Frederik H., and Patrick M. Parkinson (1986) "Recent Developments in the Bankers Acceptance Market", *Federal Reserve Bulletin*, vol. 72 (January 1986), pp. 1-12.

Joannès, Francis (2008), "Les activités bancaires en Babylonie", in Koenraad Verboven, Katelijn Vandorpe and Véronique Chankowski (eds.), *Pistoi dia tèn Technèn: Bankers, Loans and Archives in the Ancient World*, Leuven: Peeters, pp. 17-30.

Kadens, Emily (2012), "The Myth of the Customary Law Merchant", *Texas Law Review*, vol. 90, pp. 1153-1206.

Keynes, John M. (1914), "War and The Financial System – August 1914", *The Economic Journal*, vol. 24, pp. 460-486.

Kindleberger, Charles K. (1973), *The World in Depression, 1929-1939*, Los Angeles: University of California Press.

King, Wilfred T. C. (1936), *History of the London Discount Market*, London: Routledge.

LaRoche, Robert K. (1993) "Bankers Acceptances", Federal Reserve Bank of Richmond *Economic Quarterly*, vol. 79 (Winter 1993), pp. 75-85.

League of Nations (1935), *Enquiry into Clearing Agreements*, Geneva: League of Nations.

Lindert, Peter H. (1969), "Key Currencies and Gold, 1900-1913", Princeton Studies in International Finance, No 24, pp. 1-85.

Luzzatto, Gino (1954), "Vi furono fiere a Venezia?", in Gino Luzzatto, *Studi di storia economica veneziana*, Padua: CEDAM, pp. 201-209.

Malynes, Gerard (1686), *Consuetudo, vel, Lex Mercatoria: Or, the Ancient Law-Merchant*, 3rd ed., London: Bassett.

Milgrom, Paul R., Douglass C. North, and Barry R. Weingast (1990), "The Role of Institutions in the Revival of Trade: The Law Merchant, Private Judges, and the Champagne Fairs", *Economics and Politics*, vol. 2, pp. 1-23.

Moshenskyi, Sergii (2008), *History of the Weksel: Bill of Exchange and Promissory Note*, Bloomington IN: Xlibris.

Mueller, Reinhold C. (1997), *The Venetian Money Market: Banks, Panics, and the Public Debt, 1200-1500*, Baltimore MD: Johns Hopkins University Press.

Murray, Daniel E. (1994), "The U.N. Convention on International Bills of Exchange and International Promissory Notes with Some Comparisons with the Former and Revised Article Three of the UCC", *University of Miami Inter-American Law Review*, vol. 25, pp. 189-225.

Niepman, Friedericke, and Tim Schmidt-Eisenlohr (2016), "No Guarantees, No Trade: How Banks Affect Export Patterns", Board of Governors of the Federal Reserve System International Finance Discussion Papers, No 1158.

Niepman, Friedericke, and Tim Schmidt-Eisenlohr (2017), "International Trade, Risk and the Role of Banks", *Journal of International Economics*, vol. 107, pp. 111-126.

Nishimura, Shizuya (1971), *The Decline of Inland Bills of Exchange in the London Money Market, 1855-1913*, Cambridge: Cambridge University Press.

Richards, Richard D. (1929), *The Early History of Banking in England*, London: King.

Roberts, Richard (2013), *Saving the City: The Great Financial Crisis of 1914*, Oxford: Oxford University Press.

Paravisini, Daniel, Veronica Rappoport, Philipp Schnabl, and Daniel Wolfenzon (2015), "Dissecting the Effect of Credit Supply on Trade: Evidence from Matched Credit-Export Data", Review of Economic Studies, vol. 82, pp. 333-359.

Quinn, Stephen, and William Roberds, "Responding to a Shadow Banking Crisis: the Lessons of 1763", *Journal of Money, Credit and Banking*, vol. 47, pp. 1149-1176.

Rogers, James S. (1995), *The Early History of the Law of Bills and Notes: A Study of the Origins of Anglo-American Commercial Law*, Cambridge: Cambridge University Press.

Santarosa, Veronica A. (2015), "Financing Long-Distance Trade: The Joint Liability Rule and Bills of Exchange in Eighteenth-Century France", *The Journal of Economic History*, vol. 75, pp. 690-719.

Scammell, William M. (1968), *The London Discount Market*, London: Elek.

Schnabel, Isabel, and Hyun Song Shin (2004), "Liquidity and Contagion: The Crisis of 1763", *Journal of the European Economic Association*, vol. 2, pp. 929-968.

Sgard, Jérôme (2015), "Global Economic Governance During the Middle Ages: The Jurisdiction of the Champagne Fairs", *International Review of Law and Economics*, vol. 42, pp. 174-184.

Sgard, Jérôme (2018), "The Simplest Model of Global Governance Ever Seen: The London Corn Market (1885-1930)", in *Oxford Handbook of Institutions, International Economic Governance and Regulation*, forthcoming.

Spalding, William F. (1933), *The London Money Market*, London: Sir I. Pitman and Sons Ltd.

Steffenburg, E. (1949), "Merchant Banking in London", in International Banking Summer School (eds.), *Current financial problems and the City of London*, London: Institute of Bankers.

Truptil, Roger J. (1936), *British Banks and the London Money Market*, London: Jonathan Cape Ltd.

Ugolini, Stefano (2017), *The Evolution of Central Banking: Theory and History*, London: Palgrave Macmillan.

Van der Wee, Herman (1963), *The Growth of the Antwerp Market and the European Economy*, The Hague: Nijhoff.

Volckart, Oliver, and Antje Mangels (1999), "Are the Roots of the Modern Lex Mercatoria Really Medieval?", *Southern Economic Journal*, vol. 65, pp. 427-450.

Warburg, Paul M. (1910), *The Discount System in Europe*, Washington DC: National Monetary Commission.

Warburg, Paul M. (1914), "The Discount System in Europe", *Proceedings of the Academy of Political Science in the City of New York*, vol. 4, pp. 129-158.